\documentclass{ieeeaccess}

\usepackage{cite} 
\usepackage[hyphens]{url}
\usepackage[colorlinks,urlcolor=blue]{hyperref}
\hypersetup{
    colorlinks=true,
    linkcolor=blue,
    citecolor=blue,
    urlcolor=blue,
    }

\usepackage{graphicx}
\usepackage{balance}

\usepackage{textcomp}
\def\BibTeX{{\rm B\kern-.05em{\sc i\kern-.025em b}\kern-.08em
    T\kern-.1667em\lower.7ex\hbox{E}\kern-.125emX}}
\usepackage[english]{babel}

\usepackage{subcaption}
\usepackage[labelsep=period]{caption}
\usepackage[exponent-product=\cdot]{siunitx}

\usepackage{float}

\usepackage{tabularx}
\usepackage{booktabs}

\usepackage[nolist,nohyperlinks]{acronym}

\usepackage[final]{microtype}

\usepackage[intlimits]{amsmath}

\usepackage[nameinlink,capitalize]{cleveref}

\Crefname{figure}{Fig.}{Figs.}
\Crefname{table}{Table}{Tables}
\DeclareCaptionFormat{figcapfont}{\figcapfont \textbf{\textcolor{accessblue}{#1#2}}\textsf{#3}}
\def\subfigcapfont{\rmfamily\fontencoding{T1}\fontseries{n}\fontsize{8}{9.6}\selectfont} 
\DeclareCaptionFormat{subfigcapfont}{\subfigcapfont #1#2#3}
\renewcommand{\thetable}{\Roman{table}}

\begin{document}
    \begin{acronym}
        \acro{surface_distance}[ASSD]{average symmetric surface distance}
        \acro{casenet}[CASENet]{Deep Category-Aware Semantic Edge Detection}
        \acro{chrnet}[CHRNet]{Cascaded and High-Resolution Network}
        \acro{csd}[CSD]{charge stability diagram}
        \acro{ct}[CT]{charge transition}
        \acro{cnn}[CNN]{convolutional neural network}
        \acro{dff}[DFF]{Dynamic Feature Fusion}
        \acro{dice}[DICE]{Dice similarity coefficient}
        \acro{diffusionedge}[DiffusionEdge]{Diffusion Probabilistic Model for Crisp Edge Detection}
        \acro{dpm}[DPM]{diffusion probabilistic model}
        \acro{dqd}[DQD]{double quantum dot}
        \acro{ed}[ED]{Edge Drawing}
        \acro{edter}[EDTER]{Edge Detection TransformER}
        \acro{ffn}[FFN]{feedforward neural network}
        \acro{flop}[FLOP]{floating point operation}
        \acro{fpn}[FPN]{Feature Pyramid Network}
        \acro{gcanny}[GCanny]{Generalized Canny}
        \acro{gpb}[gPb]{globalized probability of boundary}
        \acro{hpo}[HPO]{hyper-parameter optimization}
        \acro{idt}[IDT]{interdot transition}
        \acro{ldt}[LDT]{lead-to-dot transition}
        \acro{ldc}[LDC]{Lightweight Dense Convolutional Neural Network}
        \acro{mac}[MAC]{multiply-accumulate operation}
        \acro{manet}[MA-Net]{Multi-Scale Attention Network}
        \acro{medsegdiff}[MedSegDiff]{Medical Image Segmentation with Diffusion Probabilistic Model}
        \acro{ml}[ML]{machine learning}
        \acro{mlp}[MLP]{multilayer perceptron}
        \acro{mmvitseg}[MMViT-Seg]{Mini-Mobile Vision Transformer for Segmentation}
        \acro{nas}[NAS]{neural architecture search}
        \acro{phcon}[PhCon]{phase congruency}
        \acro{pit}[PIT]{physics-informed tuning}
        \acro{qd}[QD]{quantum dot}
        \acro{qpc}[QPC]{quantum point contact}
        \acro{qubit}[qubit]{quantum bit}
        \acro{relu}[ReLU]{rectified linear unit}
        \acro{rbc}[RBC]{ray-based classification}
        \acro{surface_dice}[S-DICE]{normalized surface Dice}
        \acro{tct}[TCT]{total charge transition}
        \acro{teed}[TEED]{Tiny and Efficient Edge Detector}
        \acro{vit}[ViT]{vision transformer}
        \acro{vmamba}[VMamba]{visual state space model}
        \acro{vmunet}[VM-UNet]{Vision Mamba UNet}
        \acro{resnet}[ResNet]{Residual Neural Network}
    \end{acronym}

    \doi{-}
    
    \title{Automated Charge Transition Detection in Quantum Dot Charge Stability Diagrams}
   
    \author{\uppercase{Fabian~Hader}\authorrefmark{1},
            \uppercase{Fabian~Fuchs}\authorrefmark{1},
            \uppercase{Sarah~Fleitmann}\authorrefmark{1},
            \uppercase{Karin~Havemann}\authorrefmark{1},
            \uppercase{Benedikt~Scherer}\authorrefmark{1},
            \uppercase{Jan~Vogelbruch}\authorrefmark{1},
            \uppercase{Lotte~Geck}\authorrefmark{1,2},
            \uppercase{and Stefan~van~Waasen}\authorrefmark{1,3}}
    \address[1]{Peter Grünberg Institute -- Integrated Computing Architectures (ICA / PGI-4), Forschungszentrum Jülich GmbH, 52425 Jülich, Germany}
    \address[2]{System Engineering for Quantum Computing, Faculty of Electrical Engineering and Information Technology, RWTH Aachen University, 52062 Aachen}
    \address[3]{Faculty of Engineering -- Communication Systems, University of Duisburg-Essen, 47057 Duisburg, Germany}

    \markboth
    {Hader \headeretal: Automated Charge Transition Detection in Quantum Dot Charge Stability Diagrams}
    {Hader \headeretal: Automated Charge Transition Detection in Quantum Dot Charge Stability Diagrams}

    \corresp{Corresponding author: Fabian Hader (email: f.hader@fz-juelich.de).}

    \begin{abstract}
    Gate-defined semiconductor quantum dots require an appropriate number of electrons to function as qubits. The number of electrons is usually tuned by analyzing \aclp{csd}, in which charge transitions manifest as edges. Therefore, to fully automate qubit tuning, it is necessary to recognize these edges automatically and reliably. This paper investigates possible detection methods, describes their training with simulated data from the SimCATS framework, and performs a quantitative comparison with a future hardware implementation in mind. Furthermore, we investigated the quality of the optimized approaches on experimentally measured data from a GaAs and a SiGe qubit sample.
    \end{abstract}
    
    \begin{keywords}
    semiconductor quantum dots, automated tuning, charge stability diagram, quantum computing
    \end{keywords}

    \titlepgskip=-15pt
    \maketitle

    \section{Introduction}
    \label{sec:introduction}

    \PARstart{T}{uning} the number of electrons in quantum dots is essential for creating gate-defined semiconductor \acp{qubit}. 2D \acp{csd} reveal a change in the number of electrons, hereafter \ac{ct}, as an edge in the pixel information (usually floating-point representation). They can be recorded, for example, by using the conductance change of a nearby electrostatically coupled sensor dot or a \ac{qpc} \cite{elzerman_csd_qpc}. 
    The \acp{ct} must be detected robustly to realize complete automation of the tuning process. However, considering scalability, the complexity of the approaches should be minimized. Ultimately, we propose that automated tuning should be integrated into the cryostat to reduce the wiring problem (a.o. limited bandwidth, heat dissipation) originating from the connection of room temperature electronics into the cryostat \cite{geck_2019, ruffino_cryo-cmos_2022}. If it is impossible to embed all parts of the tuning (e.g., space and dissipated heat limitations), primarily integrating the initial data processing steps can reduce the amount of data to be transmitted. In particular, knowing only the binary edge information for individual pixels is sufficient for analyzing charge transitions.
    Therefore, we consider this step an essential candidate for cryostatic hardware implementation and investigate possible detection approaches, including classical and \ac{ml} methods. We used the simulation framework SimCATS \cite{simcats} to generate training and test data. In addition, we analyzed the applicability of the selected approaches to experimental data.
    
    We organized the paper as follows: First, we comprehensively introduce state-of-the-art tuning approaches that use \ac{ct} detection (\cref{sec:background}). Then, we describe the applied datasets (\cref{sec:datasets}) and the metrics and methods selected for the evaluation (\cref{sec:metrics}). Next, we describe the selected detection approaches (\cref{sec:detectors}) and their training (\cref{sec:training}). Then, we evaluate their detection quality and, for the \ac{ml} candidates, the number of parameters and speed (\cref{sec:evaluation}). Finally, we summarize the study and draw a conclusion comprising potential improvements (\cref{sec:conclusion}).

    \section{Background}
    \label{sec:background}

    Before tuning the number of electrons in \acp{qd}, automated tuning approaches adjust the \ac{qd} device to a stable global configuration of known topology in the state space with a known number of charge islands \cite{zwolak_colloquium}. In our case, we form a \ac{dqd} and tune the number of charges in each dot. A common strategy is to empty the \acp{qd} (unloading phase) and then reload the desired number of electrons on each (reloading phase). This procedure requires identifying \acp{ct} in the \acp{csd}, as the exact number of charges cannot be sensed directly \cite{zwolak_colloquium}. Subsequently, fine-tuning of couplings between multiple dots is typically performed. Different authors used classical image processing and different kinds of \ac{ml}-methods for these tasks.

    Approaches to tuning qubits to a stable known topology comprise classical \ac{ml}-methods and deep learning methods. \cite{darulova_2019} used fitting procedures and compared different classical classifiers trained on simulated and experimental data. The cross-architecture tuning solution using AI (CATSAI) \cite{severin_2024} uses a Gaussian process model of the gate voltage hypersurface and a random classifier iteratively to tune multiple parameters at once. The proposed approach was demonstrated on three device architectures and material systems. Deep learning methods for this task use proprietary \acp{cnn} \cite{kalantre_2019a, zwolak_2020, ziegler_2021a,muto_2024} or the AlexNet model \cite{liu_2022i}. The networks are either trained on simulated data only \cite{kalantre_2019a, zwolak_2020, ziegler_2021a,muto_2024} or a mix of experimental and simulated data \cite{liu_2022i}.

    First approaches to tuning quantum dots to a specific charge regime involved classical image processing. \cite{baart_2016} used template matching (Gabor filter) to identify the most bottom-left \acp{ct} and set voltages slightly above to enter the single-electron regime. For single \acp{qd}, \cite{lapointe-major_2020} suggested using a modified Hough transform or the EDLines algorithm. Classical \ac{ml}-methods were proposed by \cite{moon_2020, vanstraaten_2022}. \cite{moon_2020} used a probabilistic \ac{ml}-model in an iterative and two-stage manner to identify candidate locations on the gate voltage hypersurface, measure the data therein, and evaluate the transport features. In contrast, \cite{vanstraaten_2022} moved to hypervolumes and performed a hypothesis (Kolmogorov-Smirnov) test if the volume contains only noise. If not, a random walk combined with a score function is employed to search for \ac{dqd} features near hypervolumes. The deep learning approaches for charge regime tuning mainly differ in \ac{cnn} models, training data, and data dimensionality. \cite{durrer_2020} used several different CNNs to predict the presence of \acp{ct} in the unloading phase and the presence and orientation for the reloading phase of a GaAs triple-QD device operated in \ac{dqd} mode. The work of \cite{czischek_2022} classifies pre-processed \ac{csd} patches (5$\times$5-pixel) via an extremely small \ac{ffn} to predict the presence of \acp{ct} and to enable a future network in memristor arrays. The model was trained on synthetic data and robustly transferred to experimental data. The \ac{pit} proposed by \cite{ziegler_2023} uses rays rather than two-dimensional measurements. The method combines the \acp{rbc} classifier \cite{zwolak_2021} or a \ac{cnn} \cite{zwolak_2018} with physics knowledge to first navigate to the target global state and then performs ray measurements to tune to the target charge state.

    Fine-tuning tunnel couplings also includes proposed methods from all categories. \cite{vandiepen_2018} fitted the interdot transition sensor signals to a classical anti-crossing model. \cite{mills_2019} established virtual gates before using the Hough line transform and template matching for interdot coupling tuning. Similarly, \cite{monical_2020} used a generalization of the Hough transform (Hough anticrossing transform) but detected the locations and inclinations of possible triple points. \cite{teske_2019} proposed a classical \ac{ml}-method by combining Bayesian statistics with adapted Kalman filtering. The deep learning approach of \cite{vanesbroeck_2020} realizes an unsupervised generative model that employs variational autoencoders, consisting of an encoder and decoder both embodied in a neural network.

    The work of \cite{liu_2022i, schuff_2024a} comprises all of the above steps \cite{liu_2022i} or claims to perform fully autonomous tuning to Rabi oscillations \cite{schuff_2024a}. Both methods use deep learning or Bayesian optimization and computer vision techniques in the case of \cite{schuff_2024a}. Deep learning has also been used to perform autonomous measurements \cite{lennon_2019b, nguyen_2021b} or single-shot readouts of charge and spin states \cite{matsumoto_2021}.
    
    \section{Datasets}
    \label{sec:datasets}

    We generated datasets\footnote{Using our python package \texttt{SimCATS-Datasets}\cite{simcats_datasets_github}.} for training and evaluating edge detection approaches using our simulation framework SimCATS \cite{simcats} and used hand-labeled experimental data from the Quantum Technology Group of RWTH Aachen\footnote{\url{https://www.quantuminfo.physik.rwth-aachen.de/cms/quantuminfo/forschung/~xwpl/quantum-technology-group/}} to assess the generalization ability.

    Some neural network architectures require image resolutions divisible by a power of 2, e.g., because the resolution is halved in each encoder step and doubled in the decoder. With at most five such steps for the investigated models, we required a resolution divisible by 32. Since the initially available experimental data from a GaAs sample\footnote{Similar to the one described in \cite{volk_loading_2019}.} with a \ac{qd} employed as sensor dot have a resolution of 100$\times$100 pixels (30mV $\times$ 30mV), we chose the closest resolution of 96$\times$96 pixels (28.8mV $\times$ 28.8mV) to be able to test the networks later on experimental data\footnote{The experimental GaAs data were reduced by removing the first and last two rows and columns.}. 
    
\begin{figure}[!t]
        \centering
        \includegraphics[width=\linewidth]{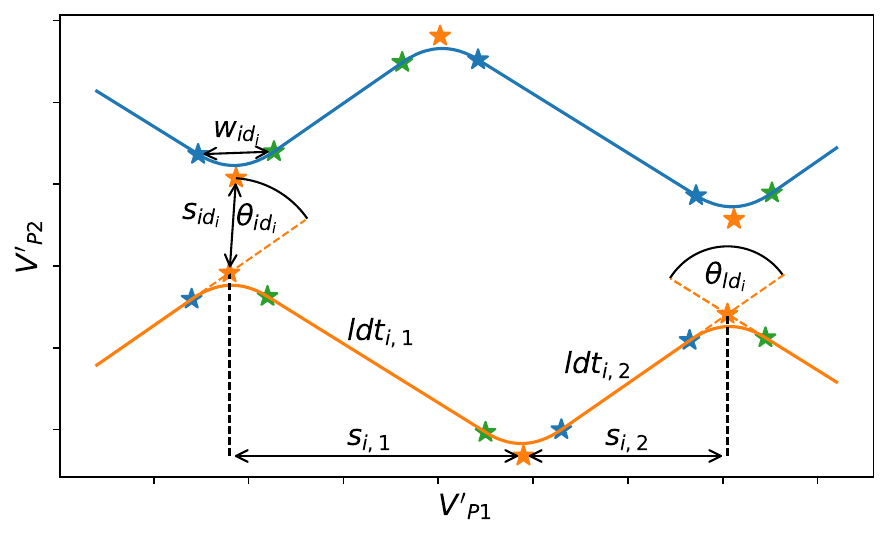}
        \caption{Visualization of parameters used to define the \acp{tct} for the geometric simulation in SimCATS.}
        \label{fig:simcats_tct_params}
    \end{figure}

    We obtained parameter ranges for the SimCATS simulations from the data of the GaAs sample as described in \cite{simcats}. Furthermore, we aimed for a diverse dataset to improve the networks' generalizability. Therefore, the generation procedure randomly selected all model parameters from the extracted ranges listed in \cref{tab:simulation_params}. While sampling most values from a uniform distribution, some parameters that describe the structure of the \acp{tct}\footnote{\Acp{tct} represent the borders between regions of the \ac{csd} containing a fixed number of electrons in the system. The \(i\)th \ac{tct} separates the regions containing \(i - 1\) and \(i\) electrons.} were sampled from a normal distribution because this matches our observations in the experimental data. In addition to the parameters described in \cite{simcats}, we supply the following parameters for the \acp{ldt} and \acp{idt}\footnote{A \acl{ldt} describes a transition between the dot system and the leads and an \acl{idt} describes the tunneling of an electron between two dots.} of the \(i\)th \ac{tct}:
    \begin{itemize}
        \item the relation between the slopes of the \acl{ldt} of dot 1 \(ldt_{i,1}\) and the \acl{ldt} of dot 2 \(ldt_{i,2}\) (\(\theta_{ld_{i}}\)), 
        \item the relation between the slopes of the \acl{idt} \(idt_{i}\) and \(ldt_{i,1}\) (\(\theta_{id_{i}}\)),
        \item the length of \(idt_{i}\) (\(s_{id_{i}}\)), 
        \item the width of \(idt_{i}\) (\(w_{id_{i}}\))\footnote{The width of \(idt_{i}\) defines the rounding at the triple points, which is controlled by the distance between the Bézier anchors \(b_{i,j}, (j{=}1,2)\) (see \cite[section~III.A.]{simcats}).}, and
        \item the relation between \(s_{id_{i}}\) and \(w_{id_{i}}\).
    \end{itemize}
    \Cref{fig:simcats_tct_params} visualizes the various parameters of the used \ac{tct} model, with the \acp{tct} displayed in the 45\(^{\circ}\)-rotated voltage space, which is denoted as \((V'_{P1}, V'_{P2})\) (see \cite[section~III.A.]{simcats}).
    In anticipation of future improvements in sample quality, we decided to include lower noise levels more often than the very high levels that are sometimes observed in the measurements. Therefore, we used an exponential distribution for these values. 
    
    For normal distributions, we set \(\mu\) to the center and \(\sigma\) to a sixth of the range to best represent the distribution in the interval. Furthermore, we selected the scale for the exponential rate to reach the 99\%{} quantile at 60\%{} of the sampling range. Generally, we resampled parameters outside the given ranges. 
    
    We do not require that the sampled parameters are physically plausible, as they differ between different samples and experimental observations are not necessarily interpretable. Thus, we expect a more diverse dataset to lead to better generalization, which also leads to a better technology independence for the trained models.

    \begin{table*}[ht]
        \caption{Parameter ranges for the simulated datasets and their distributions (uniform \(U(a,b)\), normal \(\mathcal{N}(\mu,\,\sigma^{2})\) or exponential \(\exp(\lambda)\)).
        Table rows that do not include information about a distribution are only used for a validity check of the relation of sampled parameters. The parameter names in parentheses refer to the variables used in the SimCATS paper \cite{simcats} and in \cref{fig:simcats_tct_params}. \ac{tct} parameters are given in the rotated voltage space \((V'_{P1}, V'_{P2})\). In addition to the distortions mentioned here, we also applied dot jumps as occupation distortion and random telegraph noise as sensor potential and sensor response distortion, using the parameters from the original SimCATS \texttt{default\_{}configs["GaAs\_{}v1"]} \cite{simcats_github}}.
        \centering
        \label{tab:simulation_params}
        \begin{tabularx}{\linewidth}{lXS[round-mode=figures, round-precision=2, exponent-mode=scientific, table-format=1e1, print-zero-exponent=true, print-unity-mantissa=false]S[round-mode=figures, round-precision=2, exponent-mode=scientific, table-format=1e1, print-zero-exponent=false, print-unity-mantissa=false]l}
            \hline
            & \textbf{Parameter} & \textbf{Minimum} & \textbf{Maximum} & \textbf{Distribution} \\
            \hline
            \textbf{\Aclp{tct}} & \(V'_{P1}\)-intercept of \(ldt_{i,j}, (j{=}1,2)\) (\(s_{i,j}\)) [V] & 0.01 & 0.025 & \(\mathcal{N}(\mu,\,\sigma^{2})\) \\
            & Slope of \(ldt_{i,1}\) (\(m_{i,1}\)) & -0.44 & -0.08 & \(\mathcal{N}(\mu,\,\sigma^{2})\) \\
            & Slope of \(ldt_{i,2}\) (\(m_{i,2}\)) & 0.21 & 0.55 & \(\mathcal{N}(\mu,\,\sigma^{2})\) \\
            & Angle between \(ldt_{i,1}\) and \(ldt_{i,2}\) (\(\theta_{ld_{i}}\)) [rad] & 0.43760867228 & 1.6789624091 & \\
            & Angle between \(idt_{i}\) and \(ldt_{i, 2}\) (\(\theta_{id_{i}}\)) [rad] & 0.58153916891 & 1.396938844 & \(U(a,b)\) \\
            & Length of \(idt_{i}\) (\(s_{id_{i}}\)) [V] & 0.00261 & 0.00987 & \(U(a,b)\) \\
            & Width of \(idt_{i}\) (\(w_{id_{i}}\)) [V] & 0.00043 & 0.00814 & \(U(a,b)\) \\
            & Relation of \(idt_{i}\) length to \(w_{id_{i}}\) & 0.8692269873603532 & 9.055385138137407 & \\
            \hline
            \textbf{Sensor} & Number of Coulomb peaks & \text{3} & \text{6} & \(U(a,b)\) \\
            & Lorentzian scaling factor influencing the height of the Coulomb peaks (\(a\)) & 0.02245 & 0.19204 & \(\exp(\lambda)\) \\
            & Lorentzian width influencing the width of the Coulomb peaks (\(\gamma\)) & 0.0009636 & 0.0029509 & \(U(a,b)\) \\
            & Lever arm dot 1, coupling between dot 1 and the sensor dot (\(\alpha_{1}\)) & -0.0007994 & -0.0000961 & \(U(a,b)\) \\
            & Lever arm dot 2, coupling between dot 2 and the sensor dot (\(\alpha_{2}\)) & -0.0005214 & -0.0000630 & \(U(a,b)\) \\
            & Lever arm gate 1, coupling between plunger gate 1 and the sensor dot (\(\beta_{1}\)) & 0.02805 & 0.15093 & \(U(a,b)\) \\
            & Lever arm gate 2, coupling between plunger gate 2 and the sensor dot (\(\beta_{2}\)) & 0.014025 & 0.30186 & \(U(a,b)\) \\
            & Relation of \(\beta_{1}\) to \(\beta_{2}\) & 0.5 & 2 & \\
            \hline
            \textbf{Occupation distortions} & Fermi-Dirac transition blurring \texttt{sigma} & 7.5e-5 & 6e-4 & \(\mathcal{N}(\mu,\,\sigma^{2})\) \\
            \hline
            \textbf{Sensor potential distortions} & Pink noise \texttt{sigma} & 1e-10 & 0.0005 & \(\exp(\lambda)\) \\
            \hline
            \textbf{Sensor response distortions} 
            & White noise \texttt{sigma} & 1e-10 & 0.0005 & \(\exp(\lambda)\) \\
            \hline
        \end{tabularx}
    \end{table*}
    
    The simulated train/validation/test dataset featured 10,000/1,000/1,000 different \ac{tct} and sensor configurations, each with 100 \acp{csd} generated with randomly sampled distortion strengths.
    
    The two experimental test datasets consisted of 439 (plunger, plunger)-GaAs-\acp{csd} and 81 (plunger, barrier)-SiGe{\footnote{Sample similar as described in \cite{struck_spin-epr-pair_2024}.}-\acp{csd}. As visible in \cref{fig:dataset_examples} the GaAs sample data show \ac{dqd} features, and the SiGe data single \ac{qd} features. They have a resolution of 96$\times$96 pixels (100mV $\times$ 100mV and 150mV $\times$ 150mV). We selected the SiGe voltage ranges to achieve a good balance between microscopic and macroscopic features compared with the GaAs data, although we still observed different properties. Specifically, the \acp{ct} are more blurred, and the spacing between \acp{ct} is lower, leading to more \acp{ct} per image. Nearly all SiGe-\acp{csd} featured visible \acp{ct}, as the Quantum Technology Group of RWTH Aachen specifically recorded them for our evaluation.
    
    While the ground truth \ac{ct} masks for the simulated datasets are available automatically, we manually generated masks for the experimental data. Therefore, five researchers labeled all images manually, and we combined the labels into a single binary mask afterward to improve the representation of weak lines. Notably, the SiGe data were more difficult to label, which resulted in broader line segments for the combined manual labels (see \cref{fig:dataset_examples}).

    \begin{figure*}[ht]
        \centering
        \includegraphics[width=\linewidth]{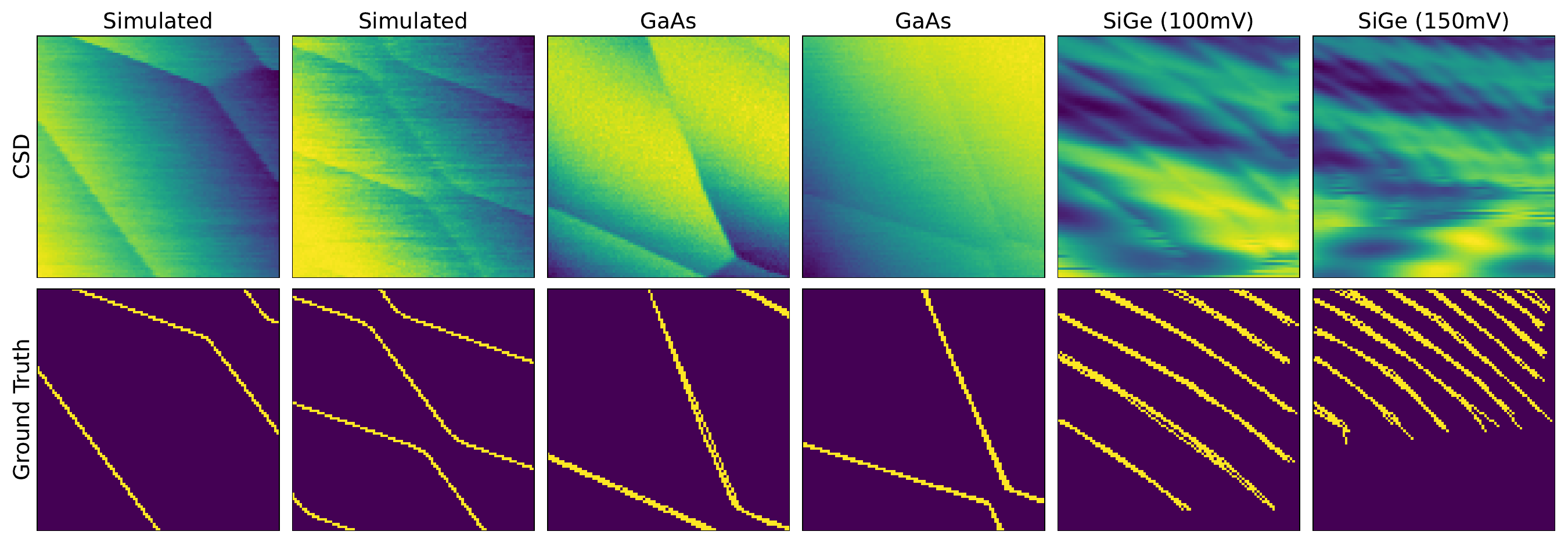}
        \caption{Examples for \acp{csd} of the final evaluation datasets and their corresponding ground truth masks.}
        \label{fig:dataset_examples}
    \end{figure*}

    \section{Metrics and Evaluation Methods}
    \label{sec:metrics}

    We used the \ac{dice} to compare the results of the different approaches and assess their quality: \[DICE=\frac{2|X\cap{}Y|}{|X|+|Y|}.\]
    This metric measures the similarity between the predicted segmentation mask \(X\) and the ground truth segmentation mask \(Y\). In our implementation, we set the \ac{dice} score to 1 in case both masks are empty. The disadvantage of the \ac{dice} score is that it expects pixel-precise segmentation and severely punishes even misalignments of one pixel. Due to the manual labeling, we did not expect pixel-precise segmentation of the experimental data. The \ac{surface_dice} score \cite{surface_dice} solves this problem by introducing a class-specific threshold for an accepted segmentation deviation in the pixel space.
    
    In addition, we measured the inference time of the models using the CUDA API because analysis speed is a crucial factor for a scalable tuning solution. The inference time specifies the calculation time required by the neural network for prediction. It does not include the time required to transfer data into GPU memory, which is identical for all approaches. 
    
    Furthermore, we used the Python package Calflops \cite{calflops} to calculate the \acp{flop} and \acp{mac} of the networks. These values are independent of the hardware used and, therefore, offer a reasonable estimate of the feasibility of cryogenic implementation on dedicated hardware.
    Moreover, we compared the number of trainable parameters, which indicates the expected implementation size.

    \section{Edge Detection Approaches}
    \label{sec:detectors}

    \cref{tab:collected_approaches} lists the approaches collected, their publication year, and their code basis.

    \begin{table}[ht]
        \caption{Collected approaches, their category, publication year, and code basis.}
        \label{tab:collected_approaches}
        \begin{tabularx}{\linewidth}{lXrr}
            \hline
            \textbf{Detector Category} & \textbf{Detector Name} & \textbf{Year} & \textbf{Code Basis} \\
            \hline
            Convolution & CASENet & 2017 & \cite{dff_src} \\
             & CHRNet & 2023 & \cite{CHRNet_src} \\
             & DeepLabV3+ & 2018 & \cite{smp_src} \\
             & DFF & 2019 & \cite{dff_src} \\ 
             & FPN & 2016 & \cite{smp_src} \\
             & LDC & 2022 & \cite{ldc_src} \\
             & LinkNet & 2017 & \cite{smp_src} \\ 
             & TEED & 2023 & \cite{teed_src} \\
             & U-Net & 2015 & \cite{unet_src} \\
             & UNet++ & 2018 & \cite{smp_src} \\
            \hline
            Transformer & CrackFormer & 2023 & \cite{crackformer_src} \\
             & EDTER & 2022 & \cite{edter_src} \\
             & MA-Net & 2020 & \cite{smp_src} \\
             & MMViT-Seg & 2023 & \cite{mmvit-seg_src} \\
             & SegFormer & 2021 & \cite{segformer_src} \\
             & Segmenter & 2021 & \cite{segmenter_src} \\
             & Swin-Unet & 2021 & \cite{swinunet_src} \\
             & TransUNet & 2021 & \cite{transunet_src} \\
            \hline
            State-Space-Model & VM-UNet & 2024 & \cite{vm-unet_src} \\
            \hline
            Diffusion & DiffusionEdge & 2024 & \cite{diffusionedge_src} \\
             & MedSegDiff-V2 & 2023 & \cite{medsegdiff2_src} \\
            \hline
            Classical & Canny & 1986 & \cite{canny_src} \\
             & CannyPF & 2015 & \cite{cannylines_src} \\
             & ED & 2011 & \cite{edlines_src} \\
             & PhCon+GCanny & 1999 & \cite{phasecongruency_src} \\
             & gPb+GCanny & 2008 & \cite{gpb_src} \\
            \hline
        \end{tabularx}
    \end{table}

    \subsection{Classical Approaches}
    \textbf{Canny}\\
    A widely used classical edge detector is the algorithm proposed by Canny \cite{canny}.
    It computes the gradient magnitude and orientation at each pixel and applies non-maximum suppression and subsequent hysteresis thresholding to determine the edge pixels.\\\\
    \newpage\noindent%
    \textbf{CannyPF}\\
    Lu et al. \cite{cannylines} proposed a parameter-free version of the Canny edge detector (CannyPF) that was included in the CannyLines line detection method.
    This method adaptively selects thresholds based on the distribution of gradient magnitude values of the image pixels.\\\\
    \textbf{\ac{gcanny}}\\
    As other local image features may be more robust at detecting \acp{ct} than the gradient, we developed a generalized version of the Canny edge detector. It uses a general feature map to replace the gradient magnitude and an optional orientation map to replace the gradient orientation as its input.
    After non-maximum suppression, it performs an additional step to connect small gaps in the resulting image.
    These connections result from checking for specific patterns of edge and non-edge pixels within a \(4\times4\) window.
    Finally, it creates a binary edge map using either binary thresholding or hysteresis thresholding.
    The features \acl{phcon} and \acl{gpb} \cite{gpb} are used to determine the edge strength.
    
    The idea of \textbf{\ac{phcon}} rests upon the local energy model proposed by Morrone et al. \cite{localenergy}, which postulates that image features become noticeable if the Fourier components are maximally in phase.
    In this paper, we computed phase congruency using an improved algorithm proposed by Kovesi \cite{kovesiphasecongruency}, which applies log-Gabor wavelets rather than Fourier components.
    
    The \textbf{\ac{gpb}}, proposed by Arbelaez et al. \cite{gpb}, combines different local image features to create a map of boundary probabilities and improves it by considering the global feature distribution afterward.\\\\
    \textbf{\ac{ed}}\\
    \ac{ed} detects edges by determining the anchor points that are most likely edge pixels based on their gradient magnitude and then establishes links between them \cite{edgedrawing}.

    \subsection{Machine Learning Approaches}
    
    \subsubsection{Convolution-Based Approaches}
    A widely used \ac{ml} approach for image processing is using \acp{cnn}. We applied networks developed for edge detection and segmentation tasks to detect \acp{ct}.\\\\
    \textbf{\ac{chrnet}}\\
    \ac{chrnet} \cite{CHRNet} detects edges using multi-scale representations of the image while preserving the high resolution of the output map. Therefore, it concatenates the output of a convolutional block with the result of the previous block, uses batch normalization layers with an active affine parameter as an erosion operation for the homogeneous region in the image, and generates the output of the network by fusing the outputs of each block.\\\\
    \textbf{\ac{ldc}}\\
    Several approaches also focus on reducing the network size, leading to faster prediction and lower hardware requirements.
    One is the \ac{ldc} \cite{ldc}. It integrates aspects of the advanced architectures DexiNed \cite{dexined} and CATS \cite{cats} but is notably smaller due to some modifications. With approximately 0.7 million parameters, it requires less than 4\% of parameters compared to DexiNed. Despite its reduced size, \ac{ldc} achieves competitive quality compared to more complex systems.
    As DexiNed, \ac{ldc} consists of layers structured in blocks. An ablation study comparing the \ac{ldc} network with three or four blocks demonstrated that the \ac{ldc} with three blocks has approximately 75\%{} fewer parameters but delivers valuable results \cite{ldc}. For \ac{ct} detection, we trained the versions of \ac{ldc} with either three (LDC-B3) or four blocks (LDC-B4).\\\\
    \textbf{\ac{teed}}\\
    \ac{teed} is another lightweight \ac{cnn} developed for simplicity, efficiency, and generalization \cite{teed}. According to the authors, with only 58k parameters, its size is less than 0.2\%{} of the state-of-the-art models.\\\\
    \textbf{\ac{casenet}}\\
    \ac{casenet} \cite{casenet} is a \ac{cnn} architecture based on \ac{resnet} \cite{resnet_2015} and a skip-layer architecture in which category-wise edge activations at the top convolution layer merge with the corresponding bottom layer features. It uses fixed fusion weights and bases the decision result primarily on high-level features.\\\\
    \textbf{\ac{dff}}\\
    \ac{dff} \cite{dff} bases upon \ac{casenet} but uses a feature extractor that normalizes the magnitude scales of multi-level features and adaptive fusion weights for different locations of multi-level feature maps, leading to finer edges.\\\\
    \textbf{U-Net}\\
    A task closely related to category-aware edge detection is semantic segmentation, which assigns objects in an image to different categories.
    U-Net is one of the most popular semantic segmentation networks. It is a fully \ac{cnn} architecture consisting of a contracting and an expansive path \cite{unet}.
    
    In addition to the standard U-Net, we investigated smaller versions of U-Net architectures with fewer layers and channels for convolution. Due to their relatively simple architecture, small U-Nets are exciting candidates for hardware implementation. Our tiny version (\textbf{UNet-38k}) has three encoder and decoder layers (four in the standard U-Net), begins with six output channels for the first convolution layer (64 in the standard U-Net), and uses bilinear upsampling. \\\\
    \textbf{UNet++}\\
    UNet++ \cite{unet++} is a more advanced version of U-Net, that uses deep supervision to segment medical images. The main difference to U-Net is the use of nested and dense skip connections to reduce the semantic gap between the feature maps of the encoder and decoder.\\\\
    \textbf{\ac{fpn}}\\
    \ac{fpn} constructs feature pyramids to detect objects at different scales at marginal extra cost \cite{fpn}. Therefore, it employs a top-down architecture with lateral connections to build high-level semantic feature maps at all scales.\\\\
    \textbf{DeepLabV3+}\\
    Another semantic segmentation approach is DeepLabV3+ \cite{deeplabv3+}. The proposed method combines spatial pyramid pooling with an encoder-decoder structure, resulting in refined segmentation results along object boundaries compared to the predecessor DeepLabV3 \cite{deeplabv3}. DeepLab approaches generally deploy dilated filters for ‘atrous convolutions’ and atrous spatial pyramid pooling to robustly segment objects at multiple scales \cite{deeplab}.\\\\
    \textbf{LinkNet}\\
    LinkNet \cite{linknet} aims for efficient semantic segmentation by using an 18-layer \ac{resnet} \cite{resnet_2015} as a light encoder and bypassing the input of each encoder layer to the output of the corresponding decoder. Thus, the decoder requires fewer parameters because it shares the knowledge learned by the encoder in each layer.

    \subsubsection{Transformer-Based Approaches}
    Another neural network type is the transformer, which has an underlying attention mechanism \cite{transformer}. While initially designed to process sequential data like text, a variant for image processing named \ac{vit} was developed \cite{visiontransformer}. However, some models do not directly incorporate a \ac{vit} but use special attention blocks, like \acs{manet}.\\\\
    \textbf{\ac{manet}}\\ 
    \ac{manet} \cite{manet} uses a self-attention mechanism to integrate local features with global dependencies adaptively. Therefore, position-wise and multi-scale fusion attention blocks capture the spatial dependencies between pixels in an overall view and the channel dependencies between feature maps.\\\\
    \textbf{Segmenter}\\
    Segmenter \cite{segmenter} uses a \ac{vit} as the encoder and employs two decoder variants for semantic segmentation. The decoder is either an ordinary linear layer or a novel mask transformer, which is a transformer encoder with multiple layers. The results presented later in this paper use the second option because it performs better on our validation dataset. \\\\
    \textbf{SegFormer}\\
    SegFormer \cite{segformer} uses a modified \ac{vit} as an encoder with a hierarchical structure without positional embedding. The decoder is a lightweight \ac{mlp} that aggregates information from different spatial resolution features arising from the hierarchical structure to combine local and global attention.\\\\
    \textbf{\ac{edter}}\\
    \ac{edter} \cite{edter} combines a transformer-based encoder and convolution-based decoder to extract precise, sharp object boundaries and meaningful edges. It simultaneously exploits the complete contextual information of the image and detailed local cues by operating in two stages. The first stage (EDTER-Global) uses the encoder part of a \ac{vit} with coarse image patches, and the second stage uses a modified local \ac{vit} encoder with finer image patches. Both stages use a bi-directional multi-level aggregation decoder, and a feature fusion module combines their results before being fed into a final decision head.\\\\
    \textbf{\ac{mmvitseg}}\\
    \ac{mmvitseg} \cite{mmvit-seg} is a lightweight model in which the encoder subnetwork is a two-path design that effectively captures the global dependence of image features and low-layer spatial details. Therefore, it uses convolutional and MobileViT blocks and a multi-query attention module to fuse multi-scale features from different levels in the decoder sub-network.\\\\
    \textbf{CrackFormer}\\
    CrackFormer \cite{crackformer} combines a SegNet-like \cite{segnet} encoder-decoder architecture with self-attention. It replaces all convolutional layers (except the first and last) with self-attention blocks and implements a feature fusion module that combines the features from each encoder-decoder stage using self-attention. Each self-attention block consists of two convolution layers, followed by a batch norm and a \ac{relu} activation function, with a self-attention layer in between.\\\\
    \textbf{TransUNet}\\
    TransUNet \cite{transunet} combines the general concepts of the U-Net architecture with a transformer. Although U-Net is effective in local feature detection, its ability to model long-range dependencies is weak. In contrast, transformers have an innate global self-attention mechanism but only limited localization capabilities due to insufficient low-level details. TransUNet combines these two architectures using a CNN-Transformer-Hybrid for the encoder and convolutional layers in the decoder. Thus, each of one's strengths overcomes the weaknesses of the other.\\\\
    \textbf{Swin-Unet}\\
    Swin-Unet \cite{swinunet} also combines the properties of the traditional U-Net's U-shaped architecture and skip-connections with a pure transformer encoder architecture. 
    The proposed method uses swin transformers \cite{swintransformer}, a variant of \acp{vit} \cite{visiontransformer}, for the encoder and a swin transformer-based decoder with patch-expanding layers to up-sample features during the expansive path.  

    \subsubsection{State-Space-Model-Based Approaches}
    State-space models are an alternative to transformers for processing long sequences \cite{stateSpace}.\\\\
    \textbf{\ac{vmunet}}\\
    \ac{vmunet} \cite{vm-unet} uses such a state-space model. Like Swin-Unet, it incorporates the architectural benefits of U-Net and non-convolutional layers. Instead of building upon \acp{vit} or, more specifically, swin transformers, it builds upon \acp{vmamba}, a recent architecture that combines a global receptive field with linear complexity.

    \subsubsection{Diffusion-Based Approaches}
    Diffusion models learn forward and reverse diffusion and are often used for image denoising, image inpainting, and image generation. Therefore, forward diffusion usually involves adding Gaussian noise to the original image, and reverse diffusion inverts that diffusion process, thereby reconstructing the original image.\\\\
    \textbf{\ac{diffusionedge}}\\
    \ac{diffusionedge} \cite{diffusionedge} is a \ac{dpm} for the general task of edge detection. The authors claimed that they avoided expensive computational resources and retained the final performance by applying a \ac{dpm} in the latent space. Thus, they enabled the classic cross-entropy loss to optimize parameters in the latent space. Additionally, they adapted a decoupled architecture to speed up the denoising process and proposed an adaptive Fourier filter to adjust the latent features of specific frequencies. This combination should result in very accurate and crisp edge maps.\\\\
    \textbf{\ac{medsegdiff}}\\
    \ac{medsegdiff} \cite{medsegdiff} is a \ac{dpm} for general medical image segmentation. It uses a modified U-Net with conditional encoding and a feature frequency parser in the reverse diffusion stage. The dynamic conditional strategy enables stepwise attention, and the feature frequency parser eliminates the high-frequency noise introduced by the former.
    
    MedSegDiff-V2 \cite{medsegdiff2} improves MedSegDiff \cite{medsegdiff} by introducing vision transformer mechanisms into the \ac{dpm}. It uses convolutional U-Nets as feature extractors but combines features using a transformer.

    \section{Training}
    \label{sec:training}

    Where applicable, we trained the networks using the original publication's optimizers, schedulers, loss functions, and hyperparameters. In addition, we trained the networks using a combination of the \texttt{AdamW} optimizer and the \texttt{OneCycleLR} scheduler (implemented in \texttt{torch.optim} \cite{pytorch}), which is the de facto state-of-the-art superconvergence method proposed in \cite{super_convergence}. In this case, the loss function consists of a combination of \texttt{BCEWithLogitsLoss} (implemented in \texttt{torch.nn} \cite{pytorch}) and \ac{dice} loss. Subsequently, we evaluated the training results of the networks on the validation dataset described in \cref{sec:datasets} and selected the best training checkpoint for final comparison. 
    \cref{tab:training_params} provides an overview of the hyperparameters used to train the final checkpoint for comparison. Classical approaches, except for gPb+GCanny, were optimized using differential evolution implemented in Scipy. For gPb+GCanny, we used a simple grid search because only one parameter was optimized.

    \begin{table*}[ht]
    \caption{Training hyperparameters of the collected \ac{ml} approaches, which are sorted by detector name. The optimizers and schedulers refer to the implementations in the \texttt{torch.optim} package \cite{pytorch}. For DiffusionEdge, we did not supply parameters because we used the original parameters.}
    \label{tab:training_params}
    \begin{tabularx}{\linewidth}{X|rr|lS[round-mode=figures, round-precision=2, exponent-mode=scientific, table-format=1e1, print-zero-exponent=false, print-unity-mantissa=false]S[round-mode=figures, round-precision=2, exponent-mode=scientific, table-format=1e1, print-zero-exponent=false, print-unity-mantissa=false]|l}
    \hline
    \multicolumn{1}{c|}{\textbf{Detector Name}} & \multicolumn{2}{c|}{\textbf{Training Settings}} & \multicolumn{3}{c|}{\textbf{Optimizer Settings}} & \multicolumn{1}{c}{\textbf{Scheduler Settings}}\\
     & \textbf{Epochs} & \textbf{Batch Size} & \textbf{Name} & \text{\textbf{Learning Rate}} & \text{\textbf{Weight Decay}} & \textbf{Name} \\
    \hline
    CASENet & 5 & 16 & Adam & 0.000010 & 0.000010 & OneCycleLR \\
    CHRNet & 5 & 16 & Adam & 0.000010 & 0.000010 & OneCycleLR \\
    CrackFormer & 5 & 64 & AdamW & 0.100000 & 0.001000 & OneCycleLR \\
    DeepLabV3+ & 5 & 16 & Adam & 0.000010 & 0.000010 & OneCycleLR \\
    DFF & 5 & 16 & Adam & 0.000010 & 0.000010 & OneCycleLR \\
    DiffusionEdge & \text{-} & \text{-} & \text{-} & \text{-} & \text{-} & \text{-} \\
    EDTER & 1 & 16 & AdamW & 0.000500 & 0.003000 & OneCycleLR \\
    EDTER-Global & 1 & 16 & AdamW & 0.000500 & 0.300000 & OneCycleLR \\
    FPN & 4 & 64 & AdamW & 0.100000 & 0.001000 & OneCycleLR \\
    LDC-B3 & 5 & 64 & AdamW & 0.010000 & 0.001000 & OneCycleLR \\
    LDC-B4 & 5 & 64 & AdamW & 0.010000 & 0.001000 & OneCycleLR \\
    LinkNet & 10 & 16 & Adam & 0.000010 & 0.000010 & OneCycleLR \\
    MA-Net & 5 & 16 & Adam & 0.000010 & 0.000010 & OneCycleLR \\
    MedSegDiff-V2 & 1 & 32 & AdamW & 0.000100 & 0 & LinearLR \\
    MMViT-Seg & 5 & 16 & Adam & 0.000010 & 0.000010 & OneCycleLR \\
    SegFormer & 1 & 256 & AdamW & 0.003000 & 0.300000 & OneCycleLR \\
    Segmenter & 5 & 32 & AdamW & 0.003000 & 0.003000 & OneCycleLR \\
    Swin-Unet & 10 & 256 & AdamW & 0.001000 & 0.010000 & OneCycleLR \\
    TEED & 5 & 8 & AdamW & 0.010000 & 0.002000 & OneCycleLR \\
    TransUNet & 4 & 64 & AdamW & 0.100000 & 0.001000 & OneCycleLR \\
    U-Net & 5 & 64 & AdamW & 0.200000 & 0.000100 & OneCycleLR \\
    UNet-38k & 5 & 64 & AdamW & 0.200000 & 0.000100 & OneCycleLR \\
    UNet++ & 5 & 64 & AdamW & 0.100000 & 0.001000 & OneCycleLR \\
    VM-UNet & 5 & 32 & AdamW & 0.001000 & 0.010000 & OneCycleLR \\
    \hline
    \end{tabularx}
    \end{table*}

    \section{Evaluation}
    \label{sec:evaluation}
    
    We evaluated the collected approaches from \cref{sec:detectors} on the test sets described in \cref{sec:datasets} using the methods from \cref{sec:metrics}.

    \Cref{tab:inferencetime} lists the number of parameters, the inference times, the \acp{flop}, and the \acp{mac} of all machine learning models.
    There are only four candidates with less than one million parameters. Of those, especially UNet-38k and \ac{teed} are very small, with 38,041 and 58,622 parameters, respectively. In addition, both are primarily based on convolutions and have a relatively simple network structure, which we consider beneficial for hardware implementation.
    The inference times for single images (batch size 1) significantly differed between the approach categories. The fastest approaches were from the convolution-based and the slowest from the diffusion-based category. In particular, DiffusionEdge and MedSegDiff-V2 are highly time-consuming and, therefore, are less applicable to scalable tuning solutions. Notably, tiny networks do not fully utilize the GPU and enable even more predictions when multiple such models run in parallel.
    In a scalable, fully automated tuning setup, a \ac{ct} detector could analyze \acp{csd} of multiple different \acp{qubit} at the same time. Therefore, we recorded inference times for a batch of 64 images.
    The given numbers of \acp{flop} and \acp{mac} depend on the number of parameters and the architecture. From the results in \cref{tab:inferencetime}, we see that small convolutional neural networks require fewer \acp{flop} and \acp{mac} than other architectures or convolutional networks with more parameters. 

    \begin{table*}[!htp]
    \caption{Size of the neural networks and their inference time, FLOPs, and MACs. We measured the inference times on an NVIDIA RTX A5000 but could not calculate FLOPs and MACs for all networks.}
    \label{tab:inferencetime}
    \begin{tabularx}{\linewidth}{Xl|r|rr|r|r}
    \hline
    \multicolumn{2}{c|}{\textbf{Detector}} & \multicolumn{1}{r|}{\textbf{Parameters}} & \multicolumn{2}{c|}{\textbf{Inference Time [ms]}} & \multicolumn{1}{c|}{\textbf{GFLOPs}} & \multicolumn{1}{c}{\textbf{GMACs}}\\
    \textbf{Name} & \textbf{Category} & \textbf{[million]} & \textbf{Batch Size 1} & \textbf{Batch Size 64} & \textbf{[per image]} & \textbf{[per image]} \\
    \hline
    UNet-38k & Convolution & 0.038 & 0.956 & 3.295 & 0.077 & 0.038 \\
    TEED & Convolution & 0.059 & 1.118 & 4.070 & 0.270 & 0.134 \\
    LDC-B3 & Convolution & 0.156 & 1.269 & 5.155 & 0.559 & 0.276 \\
    LDC-B4 & Convolution & 0.674 & 2.247 & 6.813 & 0.954 & 0.472 \\
    MMViT-Seg & Transformer & 1.013 & 23.810 & 66.641 & 0.721 & 0.354 \\
    CHRNet & Convolution & 1.450 & 2.545 & 19.167 & 4.085 & 2.037 \\
    SegFormer & Transformer & 4.446 & 5.157 & 15.429 & 5.897 & 2.946 \\
    CrackFormer & Transformer & 4.961 & 21.825 & 123.081 & 6.323 & 3.080 \\
    Segmenter & Transformer & 6.455 & 4.971 & 3775.265 & 14.517 & 7.235 \\
    LinkNet & Convolution & 11.658 & 2.190 & 6.360 & 1.420 & 0.706 \\
    U-Net & Convolution & 17.262 & 1.747 & 36.417 & 11.237 & 5.612 \\
    CASENet & Convolution & 21.793 & 2.409 & 23.293 & 16.220 & 8.100 \\
    DFF & Convolution & 21.799 & 2.683 & 24.545 & 16.226 & 8.103 \\
    FPN & Convolution & 23.149 & 3.514 & 20.702 & 3.805 & 1.899 \\
    UNet++ & Convolution & 26.072 & 4.637 & 25.619 & 10.298 & 5.141 \\
    Swin-Unet & Transformer & 27.154 & 5.994 & 32.387 & 2.160 & 1.074 \\
    VM-UNet & State-Space-Model & 27.424 & 7.432 & 99.503 & \text{-} & \text{-} \\
    MA-Net & Transformer & 31.777 & 4.337 & 14.145 & 4.628 & 2.309 \\
    DeepLabV3+ & Convolution & 45.663 & 6.779 & 22.873 & 7.834 & 3.906 \\
    MedSegDiff-V2 & Diffusion & 95.723 & 1726.027 & 41106.613 & \text{-} & \text{-} \\
    TransUNet & Transformer & 105.153 & 13.828 & 74.393 & 21.016 & 10.489 \\
    DiffusionEdge & Diffusion & 297.583 & 357.617 & 1246.634 & \text{-} & \text{-} \\
    EDTER-Global & Transformer & 322.386 & 16.215 & 12851.895 & 254.208 & 127.008 \\
    EDTER & Transformer & 416.964 & 37.759 & 25147.160 & 373.765 & 186.685 \\
    \hline
    \end{tabularx}
    \end{table*}

    \Cref{tab:metrics} summarizes the achieved metrics, sorted by the \ac{dice} score for simulated data and the \ac{surface_dice} score for experimental data. Additionally, \cref{fig:bar_plot_scores} provides a visualization of the metrics, sorted by the \ac{dice} score achieved on simulated data.
    The \ac{ml} approaches have a clear advantage over classical methods for all test datasets.
    Regarding the simulated data, U-Net-based architectures performed best (top 5), but in general, many approaches achieved convincing results. Remarkably, the small UNet-38k scored a \ac{dice} score above 0.9. In the group of classical methods, only \ac{phcon}+\ac{gcanny} was barely usable, scoring an \ac{surface_dice} score of 0.62. Notably, the global part of EDTER outperformed the entire algorithm. 
    Due to the necessary adaptation of the global patch size (from 16 to 8) to our small image resolution (96$\times$96 pixels), the cues of the local stage (patch size of 4) became too similar to the global ones, which resulted in loss of global information. 
    The results obtained on the GaAs dataset provide a good measure for the approaches' generalization ability from simulated to experimental data. Our simulated data originate from GaAs parameter ranges; thus, we expect similar scores for promising methods. Because of the uncertainty of manual labels, we used the \ac{surface_dice} score for comparison. Again, U-Net-based architectures scored the best (four out of the top five). DeepLabV3+ is unable to perform pixel-perfect segmentations, as its \ac{surface_dice} score is considerably better than its \ac{dice} score on the simulated data. 
    The proposed UNet-38k anew achieved good results (\ac{surface_dice} score above 0.89). Again, only \ac{phcon}+\ac{gcanny} achieved usable results for the classical approaches, which were even better than those for the simulated data.
    
    With the SiGe dataset, we analyzed the edge detection capabilities using an entirely different \ac{qubit} sample architecture and material. As expected, this dataset's mean \ac{surface_dice} score is lower, because of the fairly different feature properties described in \cref{sec:datasets}. Surprisingly, U-Net-based architectures no longer perform best; in particular, the original U-Net architecture is deteriorating. This indicates overfitting to double-dot features that are not present in the SiGe data. At the same time, Segmenter and \ac{ldc} performed more robust than their competitors. Again, we observed competitive results for the UNet-38k (\ac{surface_dice} score above 0.71), indicating that overfitting was not present, presumably because of its reduced capacity to learn more complex structures. All the classical approaches achieved poor results on the SiGe data because of blurrier transitions.

    \begin{table*}[!ht]
    \caption{Metrics calculated for all detectors on the three test datasets. We sorted the simulated data results by the mean \ac{dice} score and the experimental data results (GaAs \&{} SiGe) by the mean \ac{surface_dice} score, because the inaccuracies in the manual labels limit the significance of the \ac{dice} score. We calculate the \ac{surface_dice} score with an accepted segmentation deviation of two pixels.}
    \label{tab:metrics}
    \begin{tabularx}{\linewidth}{Xrr|Xrr|Xrr}
    \hline
    \multicolumn{3}{c|}{\textbf{Simulated Data}} & \multicolumn{3}{c|}{\textbf{GaAs Data}} & \multicolumn{3}{c}{\textbf{SiGe Data}}\\
    \textbf{Detector Name} & \textbf{DICE} & \textbf{S-DICE} & \textbf{Detector Name} & \textbf{DICE} & \textbf{S-DICE} & \textbf{Detector Name} & \textbf{DICE} & \textbf{S-DICE} \\
    \hline
    VM-UNet & 0.948611 & 0.984723 & Swin-Unet & 0.682991 & 0.935084 & Segmenter & 0.560009 & 0.860184 \\
    U-Net & 0.947980 & 0.983640 & VM-UNet & 0.680310 & 0.934989 & Swin-Unet & 0.529968 & 0.852079 \\
    UNet++ & 0.946648 & 0.981668 & U-Net & 0.684296 & 0.930579 & LDC-B3 & 0.542285 & 0.846194 \\
    Swin-Unet & 0.946016 & 0.984461 & DeepLabV3+ & 0.687865 & 0.930045 & LDC-B4 & 0.521855 & 0.827330 \\
    TransUNet & 0.945229 & 0.982025 & TransUNet & 0.679336 & 0.928166 & TransUNet & 0.485132 & 0.802789 \\
    CrackFormer & 0.942269 & 0.980714 & Segmenter & 0.686771 & 0.924734 & VM-UNet & 0.494951 & 0.793315 \\
    MA-Net & 0.937779 & 0.979485 & CrackFormer & 0.674050 & 0.921575 & UNet++ & 0.494466 & 0.789553 \\
    LinkNet & 0.935520 & 0.980958 & FPN & 0.680005 & 0.914432 & EDTER-Global & 0.533004 & 0.785397 \\
    CASENet & 0.925680 & 0.977085 & CHRNet & 0.670291 & 0.912873 & CASENet & 0.451935 & 0.761143 \\
    DFF & 0.923153 & 0.976307 & UNet++ & 0.669483 & 0.912810 & TEED & 0.465905 & 0.737797 \\
    MMViT-Seg & 0.912982 & 0.972229 & LDC-B4 & 0.659616 & 0.902324 & EDTER & 0.481658 & 0.734945 \\
    LDC-B4 & 0.912331 & 0.961914 & MMViT-Seg & 0.653716 & 0.892694 & UNet-38k & 0.426954 & 0.717193 \\
    LDC-B3 & 0.912062 & 0.961828 & UNet-38k & 0.640319 & 0.890312 & DFF & 0.434880 & 0.692505 \\
    UNet-38k & 0.907693 & 0.963606 & LDC-B3 & 0.643522 & 0.888084 & CrackFormer & 0.404251 & 0.660857 \\
    CHRNet & 0.888840 & 0.971695 & MA-Net & 0.654963 & 0.886866 & MA-Net & 0.398269 & 0.660703 \\
    MedSegDiff-V2 & 0.866172 & 0.944311 & LinkNet & 0.623341 & 0.870476 & LinkNet & 0.398087 & 0.658808 \\
    TEED & 0.850516 & 0.923503 & SegFormer & 0.571092 & 0.862221 & MMViT-Seg & 0.402314 & 0.644663 \\
    FPN & 0.775319 & 0.977872 & MedSegDiff-V2 & 0.607878 & 0.855302 & SegFormer & 0.370511 & 0.637816 \\
    Segmenter & 0.772588 & 0.975522 & DFF & 0.586459 & 0.834489 & MedSegDiff-V2 & 0.380844 & 0.633545 \\
    DeepLabV3+ & 0.768866 & 0.980459 & TEED & 0.606017 & 0.829216 & CHRNet & 0.358583 & 0.598633 \\
    EDTER-Global & 0.755435 & 0.875510 & CASENet & 0.565923 & 0.815075 & DiffusionEdge & 0.367199 & 0.594880 \\
    EDTER & 0.747603 & 0.874246 & EDTER & 0.544918 & 0.791837 & U-Net & 0.335262 & 0.593540 \\
    DiffusionEdge & 0.726628 & 0.864524 & EDTER-Global & 0.559983 & 0.777607 & DeepLabV3+ & 0.326072 & 0.576167 \\
    SegFormer & 0.500667 & 0.880487 & DiffusionEdge & 0.411255 & 0.753454 & FPN & 0.306444 & 0.535457 \\
    PhCon+GCanny & 0.317213 & 0.619396 & PhCon+GCanny & 0.419984 & 0.677735 & PhCon+GCanny & 0.223991 & 0.484963 \\
    CannyPF & 0.224394 & 0.427694 & CannyPF & 0.178615 & 0.390894 & gPb+GCanny & 0.012346 & 0.102248 \\
    Canny & 0.145899 & 0.371413 & Canny & 0.156118 & 0.385560 & ED & 0.022423 & 0.093937 \\
    ED & 0.141768 & 0.390697 & ED & 0.132090 & 0.367629 & Canny & 0.016921 & 0.081467 \\
    gPb+GCanny & 0.120203 & 0.209391 & gPb+GCanny & 0.170843 & 0.249936 & CannyPF & 0.012614 & 0.068481 \\
    \hline
    \end{tabularx}
    \end{table*}

    \begin{figure*}[ht]
        \centering
        \includegraphics[width=1\linewidth]{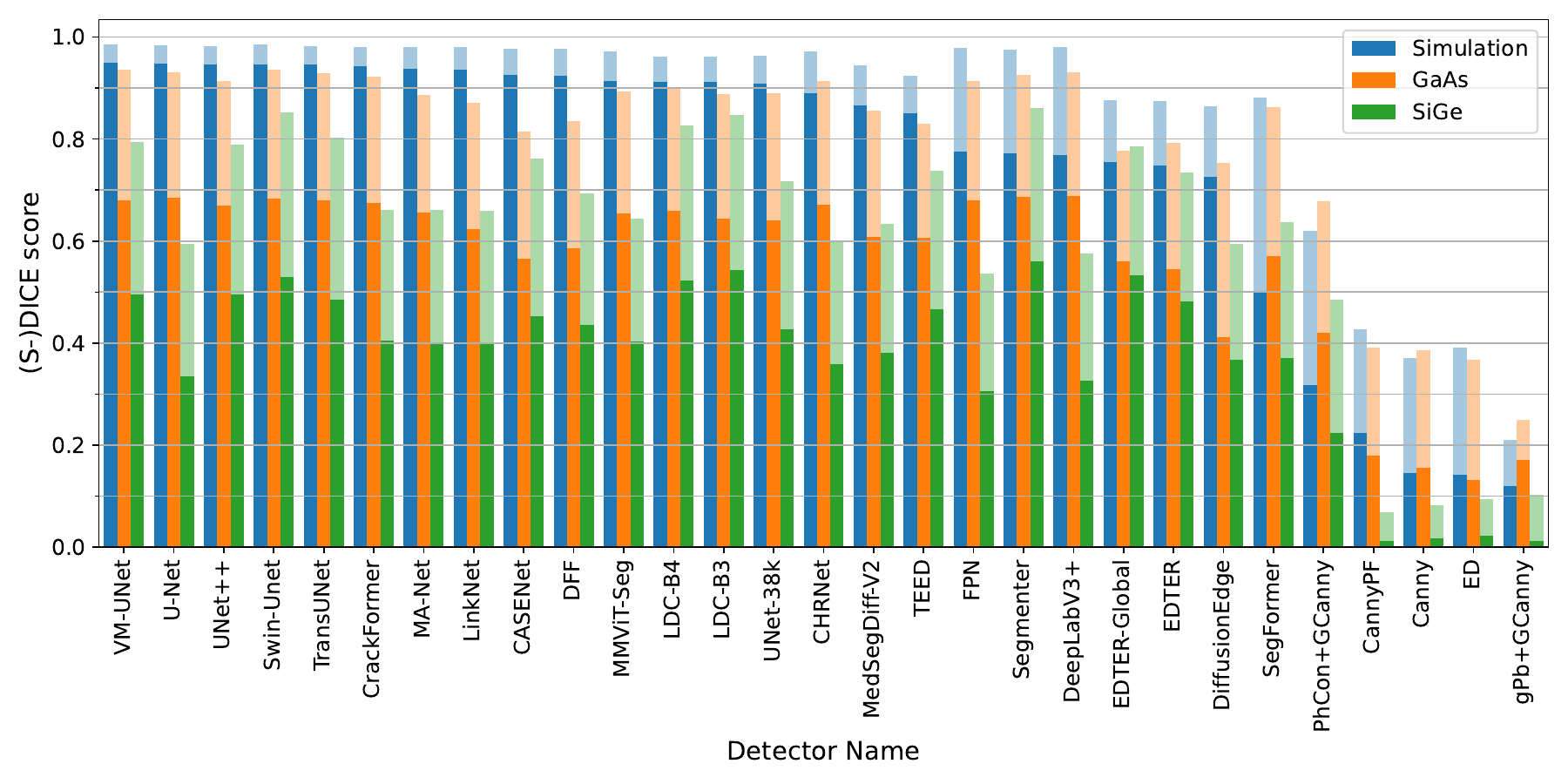}
        \caption{Bar plot visualizing the results from \cref{tab:metrics}. For each detector, the solid portion of the bar represents the \ac{dice} score, while the semi-transparent extension indicates the corresponding \ac{surface_dice} score. The detectors are arranged in accordance with their \ac{dice} score on simulation data, which is consistent with the ordering in \cref{tab:metrics}.}
        \label{fig:bar_plot_scores}
    \end{figure*}

    To summarize, we consider many approaches to have good overall analysis capabilities. Among the classical approaches, \ac{phcon}+\ac{gcanny} is the only approach that delivered valuable results. The best convolution-based approach appears to be \mbox{U-Net}, which performed poorly on the SiGe data\footnote{The poor performance on SiGe data is explained with overfitting, which can be avoided as shown with UNet-38k.}. Surprisingly robust and efficient, the scaled-down UNet-38k approach did not cause problems with the SiGe data. Swin-Unet emerged as the best approach among the analyzed transformer models. The state-space model approach VM-Unet also detected \acp{ct} robustly. Only diffusion networks are unsuitable for our application because they are too complex for energy-efficient hardware implementation and demonstrated poor metrics in our analysis. 
    \cref{fig:example_predictions} shows exemplary predictions of the best representatives of the respective approach categories. Although \ac{phcon}+\ac{gcanny} provided useful predictions for some images, it is very susceptible to distortions and the properties of the \ac{ct} features. The \ac{ml} approaches did not noticeably differ in their simulated and GaAs data predictions. Apparent differences were only visible when generalizing to the SiGe data. For example, U-Net demonstrated significant gaps in the predicted edges, whereas UNet-38k performed considerably better, and Swin-Unet worked the best.
    
    \begin{figure*}[ht]
        \centering
        \includegraphics[width=1\linewidth]{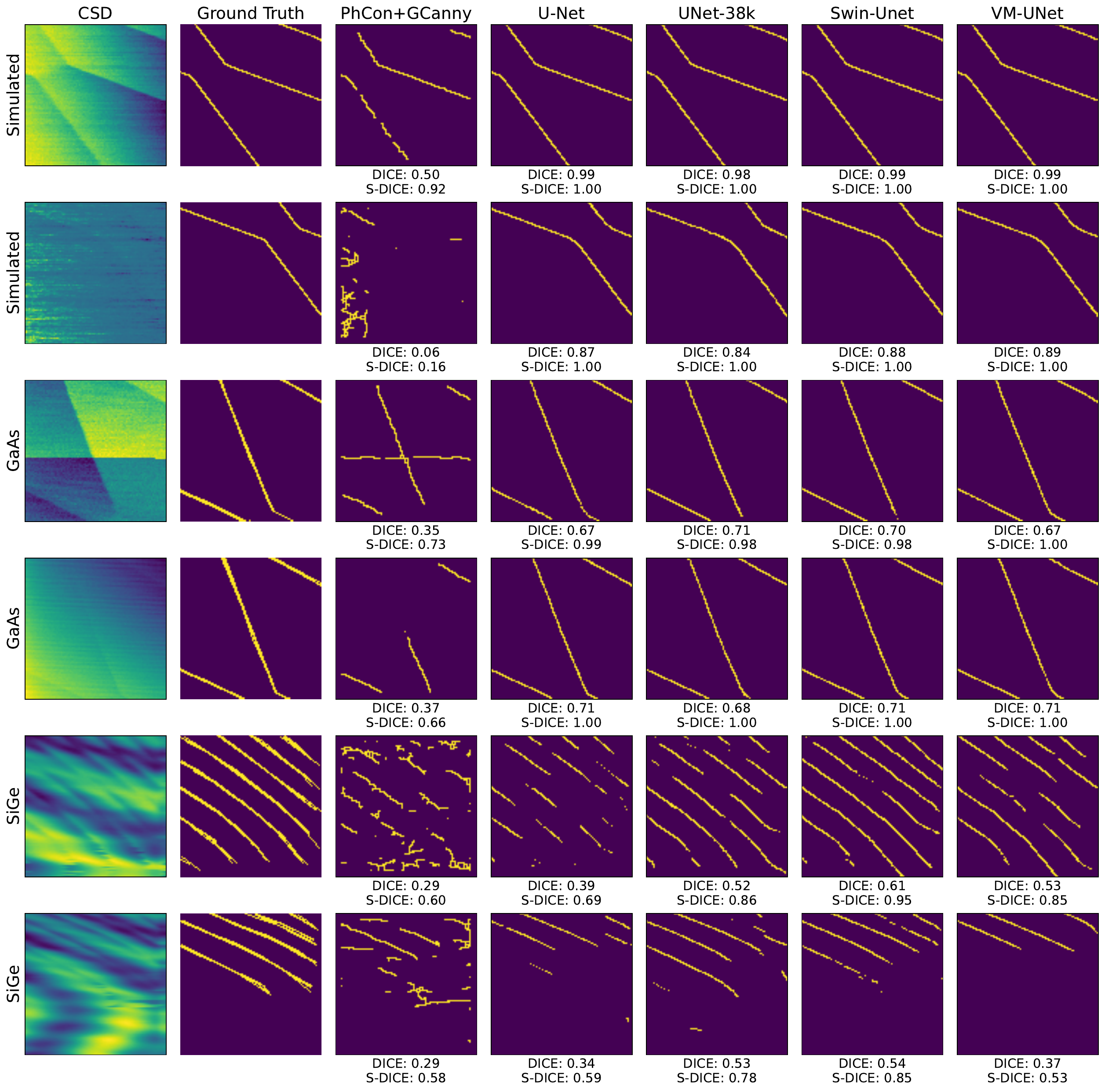}
        \caption{Exemplary predictions on the test datasets along with their corresponding \ac{dice} and \ac{surface_dice} scores. The examples were chosen to encompass a broad spectrum of cases. The top two rows present simulated data. The upper row corresponds to a sensor dot positioned on the flank of a Coulomb peak, within the regime of optimal charge sensitivity. The lower row illustrates a case where the sensor dot resides in a less favorable operating regime. The GaAs measurements display two illustrative cases: the upper panel features pronounced \acp{ct} superimposed with strong random telegraph noise, which manifests as a spurious linear feature that must not be misinterpreted as a \ac{ct}; the lower panel exhibits only faintly discernible \acp{ct}. The SiGe data show single \ac{qd} features with variations in both the angular orientation and sharpness of the \acp{ct}.}
        \label{fig:example_predictions}
    \end{figure*}

    \section{Conclusion}
    \label{sec:conclusion}

    In this study, we evaluated various classical and \ac{ml} approaches for detecting \acp{ct} in \acp{csd} using simulated data from the SimCATS framework. Our focus was on the search for suitable approach categories for robust detection with potential for future qubit-near hardware implementation.
    
    We subdivided the \ac{ml}-based approaches into \mbox{convolution-,} transformer-, state-space-model-, and diffusion-based approaches to analyze the abilities of different architectures. The analysis of the detection metrics has shown that \ac{ml}-based approaches are far superior to their classical competitors. All investigated \ac{ml} architecture categories featured valuable candidates except for the diffusion-based approaches. The results demonstrate that approaches trained on our simulated dataset can generalize to experimental data.
    
    We recommend convolution-based architectures for hardware implementation due to their low complexity and convincing detection results. Furthermore, we emphasize the possibility of creating smaller versions of networks that still have sufficient detection ability, as demonstrated with UNet-38k.    
    For experiments unconstrained by computational limitations or strict energy efficiency requirements, we recommend Swin-Unet, which consistently achieved top-tier performance across all evaluated datasets.
    \enlargethispage{\baselineskip}
    Future research should investigate further tiny versions of convolution-based networks, e.g., using \ac{nas} \cite{ren_nas_2021, benmeziane_nas_2021} and \ac{hpo} \cite{shekhar_hpo_2022} techniques. In addition to network size, a reduced data representation rather than 32-bit floating-point numbers is preferable for energy-optimized hardware implementations. The test dataset is available for benchmarking \cite{random_variation_dataset}. Furthermore, dedicated analysis hardware must prove its applicability via in-field tests at cryogenic temperatures. In this context, specialized hardware components like memristor crossbar arrays should also be considered \cite{czischek_2022}.
    
    \ifCLASSOPTIONcaptionsoff
      \newpage
    \fi

    \clearpage
    \balance
    \bibliographystyle{IEEEtran}
    \bibliography{IEEEabrv, bibliography}

    \begin{IEEEbiography}[{\includegraphics[width=1in,height=1.25in,clip,keepaspectratio]{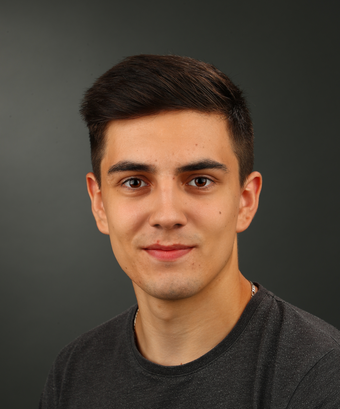}}]
    {Fabian Hader} received a B.Sc. in scientific programming and an M.Sc. in energy economics {\&} informatics from FH Aachen -  University of Applied Sciences, Jülich, Germany, in 2019 and 2021, respectively. He is pursuing a Ph.D. in engineering at the University of Duisburg-Essen, Duisburg/Essen, Germany.
    
    From 2019 to 2021, he was a Software Engineer at the Peter Grünberg Institute -- Integrated Computing Architectures (ICA / PGI-4), Forschungszentrum Jülich GmbH, Germany. His research interest focuses on the automatic tuning of \acp{qd}.
    \end{IEEEbiography}

    \begin{IEEEbiography}[{\includegraphics[width=1in,height=1.25in,clip,keepaspectratio]{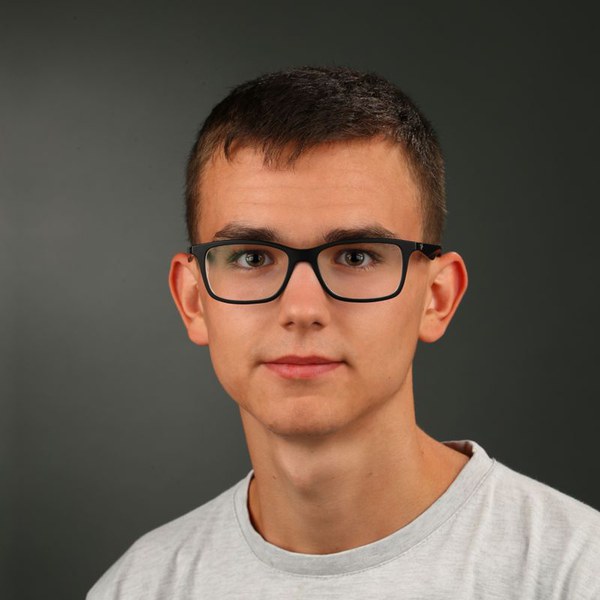}}]
    {Fabian Fuchs} received a B.Sc. in Applied Mathematics and Computer Science and an M.Sc. in Applied Mathematics and Computer Science from FH Aachen -- University of Applied Sciences, Campus Jülich, Germany in 2021 and 2023, respectively. He also obtained an M.Sc. in Mathematics from the University of Wisconsin-Milwaukee in 2023. He is pursuing a Ph.D. in Computer Science at the University of Mannheim in cooperation with the Fraunhofer Institute for Industrial Mathematics (ITWM). 
    
    From 2018 to June 2024, he worked as a software engineer at the Peter Grünberg Institute -- Integrated Computing Architectures (ICA / PGI-4), Forschungszentrum Jülich GmbH, Germany. His research interests are deep learning and computer vision.
    \end{IEEEbiography}
    
    \begin{IEEEbiography}[{\includegraphics[width=1in,height=1.25in,clip,keepaspectratio]{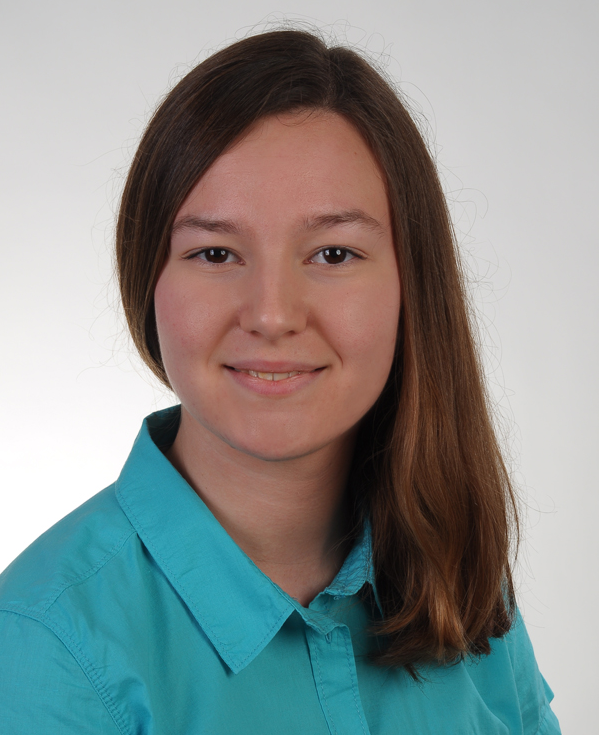}}]
    {Sarah Fleitmann} received a B.Sc. in scientific programming and an M.Sc. in applied mathematics and informatics from the FH Aachen -- University of Applied Sciences, Campus Jülich, Germany, in 2020 and 2022, respectively.\\
    Since 2017, she has been working as a software engineer at the Peter Grünberg Institute -- Integrated Computing Architectures (ICA / PGI-4), Forschungszentrum Jülich GmbH, Germany. Her research interests include the automatic tuning of quantum dots for their operation as qubits.
    \end{IEEEbiography}
    
    \clearpage
    \nobalance

    \begin{IEEEbiography}[{\includegraphics[width=1in,height=1.25in,clip,keepaspectratio]{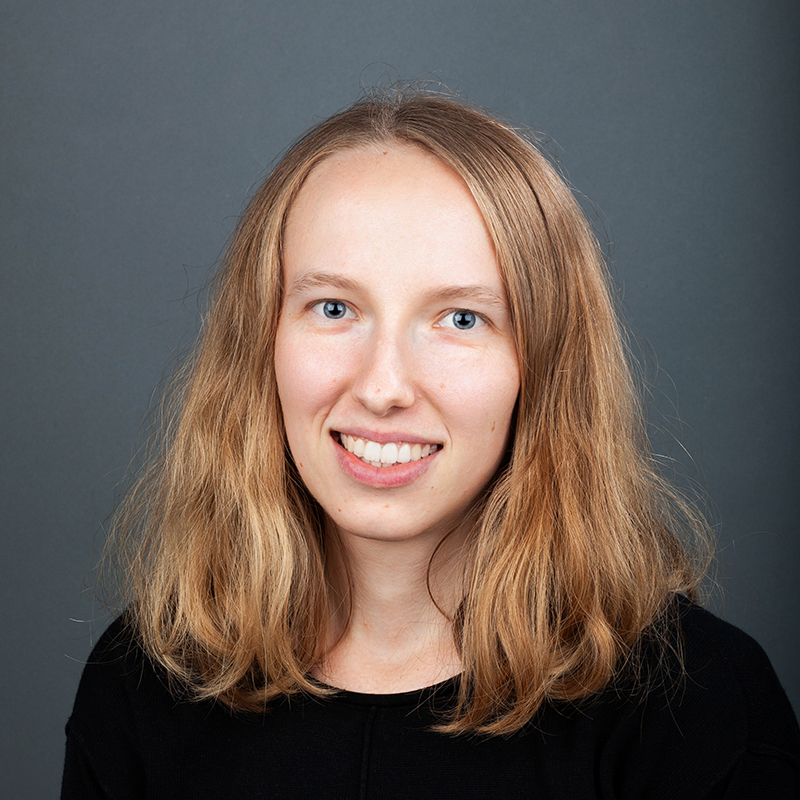}}]
    {Karin Havemann} received a B.Sc. in Applied Mathematics and Computer Science from FH Aachen -- University of Applied Sciences, Campus Jülich, Germany, in 2022. She is pursuing an M.Sc. in Applied Mathematics and Computer Science from FH Aachen -- University of Applied Sciences, Campus Jülich, Germany. Since 2023, she has worked as a software engineer at the Peter Grünberg Institute -- Integrated Computing Architectures (ICA / PGI-4), Forschungszentrum Jülich GmbH, Germany.
    \end{IEEEbiography}

    \begin{IEEEbiography}[{\includegraphics[width=1in,height=1.25in,clip,keepaspectratio]{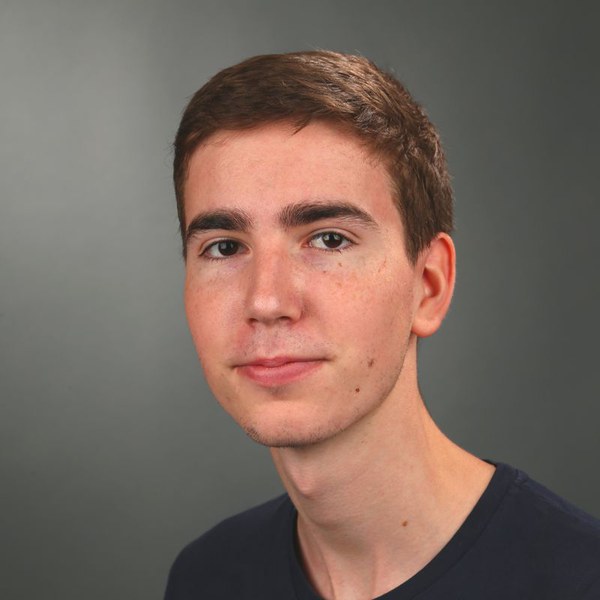}}]
    {Benedikt Scherer} received a B.Sc. in scientific programming and an M.Sc. in applied mathematics and informatics from the FH Aachen -- University of Applied Sciences, Campus Jülich, Germany, in 2020 and 2023, respectively.\\
    Since 2017, he has worked as a software engineer at the Peter Grünberg Institute -- Integrated Computing Architectures (ICA / PGI-4), Forschungszentrum Jülich GmbH, Germany.
    \end{IEEEbiography}
    
    \begin{IEEEbiography}[{\includegraphics[width=1in,height=1.25in,clip,keepaspectratio]{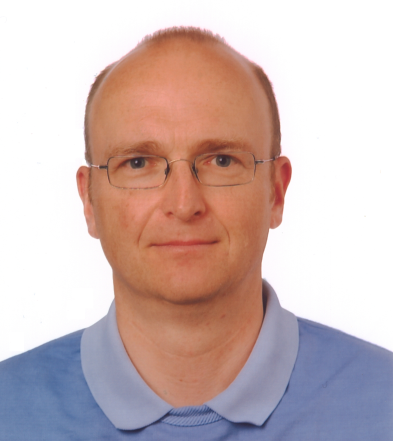}}]
    {Jan Vogelbruch} received the Dipl.Ing. and Dr.-Ing. degrees in electrical engineering from the RWTH Aachen University, Germany, in 1994 and 2003, respectively.\\
    In 1995, he joined Parsytec Computer GmbH, Aachen, Germany, as a technical project manager for European cooperations. His focus has been on high-performance computing and image processing solutions, where he has been the technical leader for the company's part in several EC-funded projects.
    Since late 1998, he has been with the Peter Grünberg Institute -- Integrated Computing Architectures (ICA / PGI-4), Forschungszentrum Jülich GmbH, Germany. His research interests include parallel computing, signal and 3D image processing, fast reconstruction methods for high-resolution computer tomography, and automated defect detection. His current research focus is on the automatic tuning of semiconductor quantum dots.
    \end{IEEEbiography}
    
    \begin{IEEEbiography}[{\includegraphics[width=1in,height=1.25in,clip,keepaspectratio]{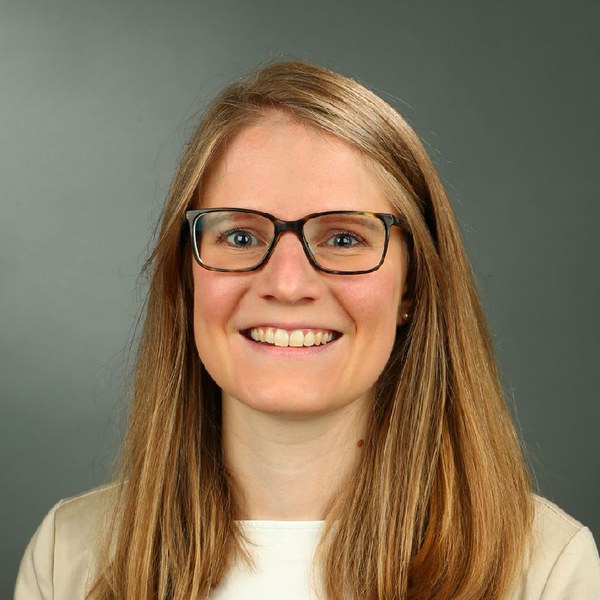}}]
    {Lotte Geck} received a B.Sc. and an M.Sc. from the RWTH Aachen University in 2013 and 2015, respectively. In 2016, she joined the Peter Grünberg Institute -- Integrated Computing Architectures (ICA / PGI-4), Forschungszentrum Jülich GmbH, Germany. She received the Dr.-Ing. degree in 2021. Since 2022, she has been a Junior Professor at Forschungszentrum Jülich and RWTH Aachen University. Her research interests include scalable electronic system solutions for quantum computing.
    \end{IEEEbiography}

    \begin{IEEEbiography}[{\includegraphics[width=1in,height=1.25in,clip,keepaspectratio]{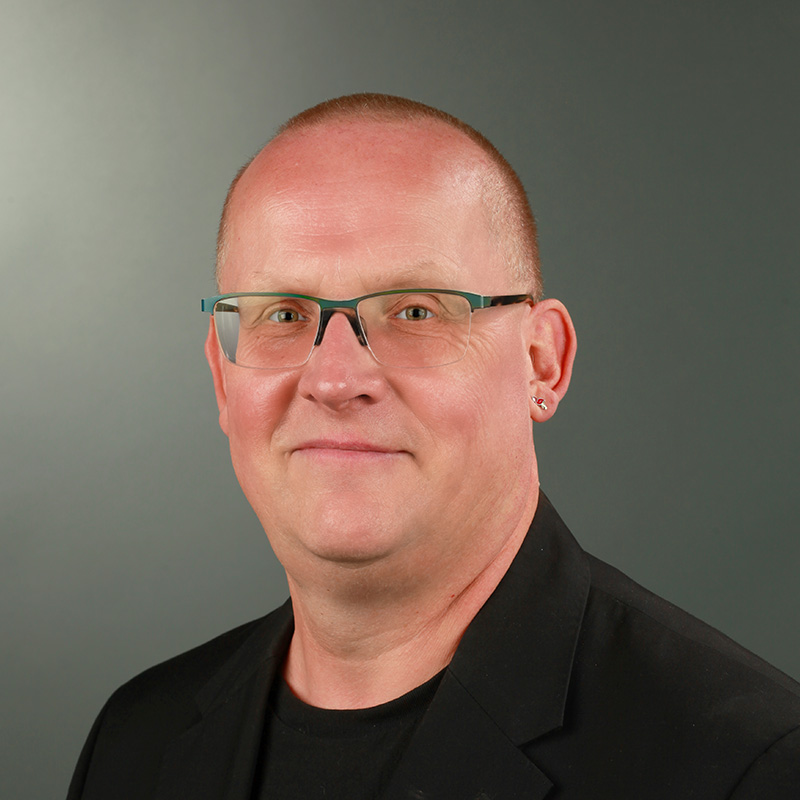}}]
    {Stefan van Waasen} received the Dipl.-Ing. and Dr.-Ing. degrees in electrical engineering from Gerhard-Mercator University, Duisburg, Germany, in 1994 and 1999, respectively. The topic of his doctoral thesis was optical receivers up to 60 Gb/s based on traveling wave amplifiers.\\
    In 1998, he joined Siemens Semiconductors/Infineon Technologies AG, Düsseldorf, Germany. His responsibility was to develop BiCMOS and CMOS RF systems for highly integrated cordless systems, such as DECT and Bluetooth. In 2001, he moved to the IC development of front-end systems for high-data-rate optical communication systems. From 2004 to 2006, he worked at the Stockholm Design Center and was responsible for the short-range analog, mixed-signal, and RF development for SoC CMOS solutions. From 2006 to 2010, he was responsible for wireless RF system engineering in the area of SoC CMOS products at headquarters in Munich, Germany, and later at the Design Center Duisburg, Duisburg. Since 2010, he has been the Director of the Peter Grünberg Institute -- Integrated Computing Architectures (ICA / PGI-4), Forschungszentrum Jülich GmbH, Germany. In 2014, he became a professor in measurement and sensor systems at the Communication Systems Chair of the University of Duisburg-Essen. His research focuses on complex measurement and detector systems, particularly on electronic systems for Quantum Computing.
    \end{IEEEbiography}

    \clearpage{}
    
    \setcounter{figure}{0} \renewcommand{\thefigure}{A.\arabic{figure}}
    \setcounter{table}{0} \renewcommand{\thetable}{A.\arabic{table}}

    \EOD

\end{document}